# Decomposing Excess Commuting: A Monte Carlo Simulation Approach[1]

Yujie Hu[1], Fahui Wang[1]
[1]Department of Geography & Anthropology, Louisiana State University, Baton Rouge, LA 70803, USA

**Abstract**

*Excess* or *wasteful* commuting is measured as the proportion of actual commute that is over minimum (optimal) commute when assuming that people could freely swap their homes and jobs in a city. Studies usually rely on survey data to define actual commute, and measure the optimal commute at an aggregate zonal level by linear programming (LP). Travel time from a survey could include reporting errors and respondents might not be representative of the areas they reside; and the derived optimal commute at an aggregate areal level is also subject to the zonal effect. Both may bias the estimate of excess commuting. Based on the 2006-2010 Census for Transportation Planning Package (CTPP) data in Baton Rouge, Louisiana, this research uses a Monte Carlo approach to simulate individual resident workers and individual jobs within census tracts, estimate commute distance and time from journey-to-work trips, and define the optimal commute based on simulated individual locations. Findings indicate that both reporting errors and the use of aggregate zonal data contribute to miscalculation of excess commuting.

**Keywords:** excess or wasteful commuting; linear programming; zonal effect; modifiable areal unit problem (MAUP); Monte Carlo simulation; CTPP

## Introduction

The past three decades have witnessed that the growth of urban travel has significantly outpaced that of population in the U.S. According to Table VM-202 in Highway Statistics (U.S. DOT/FHWA, 2012) and Table 7 in Statistical Abstract of the United States (U.S. Census Bureau, 2012), the Vehicle Miles Travelled (VMT) in the U.S. urban areas has increased by 133% from 1980 to 2012 while the population increased by 36% during the same period. Local or regional jobs-housing imbalance may help explain part of the increasing travel inefficiency (Cervero, 1989; Sultana, 2002; Sultana and Weber, 2014; Wang 2001). A closely related body of literature is *excess commuting* that also focuses on the relationship between travel efficiency and urban structure.

---

[1] This is a preprint of: Hu, Y., & Wang, F. (2015). Decomposing excess commuting: A Monte Carlo simulation approach. *Journal of Transport Geography, 44*, 43-52.
https://doi.org/10.1016/j.jtrangeo.2015.03.002

Excess commuting is a non-optimal urban work travel, measured as the proportion of the actual commute that is over the required (optimal or minimum) commute suggested by the spatial arrangement of homes and jobs in a city. Proposed first by urban economist Hamilton (1982), it has become a popular topic in economics, geography, urban planning and civil engineering, cumulating a large body of literature. However, there remain issues in both estimates of actual and optimal commutes. For instance, actual commute is usually based on survey data with reporting errors and inconsistent recall of travel time from respondents, and estimated optimal commute often varies with aggregate areal levels and thus is subject to the Modifiable Areal Unit Problem (MAUP).

Based on the 2006-2010 Census for Transportation Planning Package (CTPP) data in Baton Rouge, Louisiana, this research uses a Monte Carlo approach to simulate individual resident workers (O) and individual jobs (D) within census tracts so that it is free of the effect of areal unit size. Factual commute amount, measured in both travel time and distance, is estimated from reported O-D trips and thus establishes a reliable baseline for actual commute. By doing so, we are able to mitigate the two major contributors to the miscalculation of excess commuting in the literature.

**Background**

The concept of excess/wasteful commuting was first proposed by Hamilton (1982), as he was to examine if the classical urban economic model would perform a good job in predicting the mean commute length in a city. Hamilton assumed that both population and employment densities decline exponentially with distance from the city center, and the latter have a steeper gradient than the former. Assuming that residents could freely swap houses, the optimal (minimal) commuting pattern is that people always commute toward the city center and the trips end at the nearest jobs. As a result, the average minimal commute is the difference in average distances of population and employment from the city center. Surprisingly, he found that actual commute was about 87% excess in comparison to the optimal in 14 U.S. cities. White (1988) argued that the urban commute optimization should be constrained to the existing spatial distribution of homes and jobs and the road network, and formulated the optimal commuting pattern by a simple Linear Programming (LP) approach. White's model returned only 11% excess commuting for the same study areas used by Hamilton. The large gap in the results by Hamilton and White has led to a sustained debate on how to accurately measure excess commuting, and generated a large volume of literature from multiple disciplines.

Some attributed the discrepancy to the *scale (zonal) effect*. Hamilton (1989) cautioned researchers of assuming an optimal intrazonal commute, and pointed out that a larger zone size might lead to less excess commuting. In practice, the LP result usually yields a high proportion of optimal commute trips within a zone, e.g., 90.7% in Small and Song (1992) were intrazonal commute. If one adopts average reported intrazonal commuting time for all zones in a study area, the optimal commute is likely to be overestimated. An inflated optimal commute leads to an underestimated excess commuting such as the case for White (1988) that was only 11%. Based on a smaller zone, i.e., Traffic Analysis Zones (TAZs), Small and Song (1992) implemented White's LP approach in Los Angeles and found 66% excess commuting, substantially higher than White's but lower than Hamilton's. Their finding confirmed Hamilton's proposition of scale effect. Horner and Murray (2002) linked the issue to the Modifiable Areal Unit Problem (MAUP), which is well known to geographers (Openshaw and Taylor, 1979). They further validated the impact of spatial unit definition on the estimation of excess commuting and suggested using zonal data as disaggregate as possible.

Others suspected that different metrics might play a role (Hamilton, 1989). Hamilton (1982) used *distance* while White (1988) used *travel time*. Most found a high consistency between the two and rejected that it was a major factor causing the discrepancy (e.g., Cropper and Gordon, 1991; Small and Song, 1992).

Some believed that job decentralization might account for most or all of the excess commuting (Merriman et al., 1995; Suh, 1990). However, as argued by Giuliano and Small (1993), the direction of job decentralization's impact on commuting could be ambiguous. On the one side, it may encourage urban sprawl, reduce the land use intensity and thus increase commute lengths. On the other side, it may also improve the jobs-housing balance in many areas particularly suburbia and alleviate the need of lengthy commute to downtown. Following the suggestions of Hamilton (1982), several studies searched for factors beyond land use that prevented people from attaining optimal commute. Some emphasized the variability of labor participation rate across households as it is harder for households with multiple workers to optimize commuting for individuals (Kim, 1995; Thurston and Yezer, 1991). Certainly, residential choice is hardly made solely for the purpose of minimal commuting and often involves a complex decision considering also housing and neighborhood attributes (Cropper and Gordon, 1991). Not all workers are mobile in terms of residential choice and thus limit their likelihood of relocating to save commute (Buliung and Kanaroglou, 2002). Furthermore, the boundary effect for a study area may also affect the magnitude of excess commuting as Frost et al. (1998) found in their study of British cities. Our fascination of excess commuting is also linked to its implication in public policy (Fan et al., 2011). Many studied the issue for various transportation or land use related policy scenarios that could reduce commuting cost and their related environmental and economic impacts (Boussauw et al., 2012; Horner, 2002; Ma and Banister, 2006; O'Kelly and Lee, 2005; Rodríguez, 2004; Scott et al., 1997; Yang, 2008).

This research directs the attention back to the measurement of excess commuting given a land use pattern (i.e., employment and resident worker locations). Is it possible to design a study as disaggregate as possible to mitigate (or perhaps be even free from) the scale effect for estimating required commute, as suggested by Horner and Murray (2002)? Given the concern related to privacy, it is unlikely for researchers to access a large scale commuting dataset of individual households geocoded to their home addresses and workplaces. Our approach is to simulate individual employment and resident worker locations and the trips between them in order to alleviate the concern of unreliable estimate of intrazonal commute and also improve the accuracy of estimating interzonal commute lengths.

Another issue is on how actual commute is measured. Some commuting literature (though not directly related to excess commuting) suggested the unreliability of measuring commute length by travel time. It could be misleading as travel time by slower modes (e.g., carpool, transit, walking or bicycling) is much longer than drove-alone for the same distance. The journey-to-work survey (e.g., CTPP) data often contain some erroneous records (e.g., commute trips of several hours for travelling only a few miles). There is also concern of inconsistency in the way respondents report commute time (e.g., whether including "mental time" as noted in Wang (2003)). Excess commuting calculated at different times may also vary (Frost et al., 1998; Yang, 2008). As excess commuting is the difference between actual and optimal commuting, it is an unfair comparison to define actual commuting by reported time and optimal commuting by estimation. We propose to measure both by estimated travel time as well as distance through the road network, and identify the true extent of excess commuting.

**Study Area and Data Preparation**

This research uses the 2006-2010 Census Transportation Planning Package (CTPP) data that were made available recently (http://ctpp.transportation.org/Pages/5-Year-Data.aspx). Different from the previous (e.g., the 1990 or 2000) CTPP that was extracted from the long form census, this 5-year CTPP data set was based on the American Community Survey (ACS) 2006-2010. The data includes zone levels such as Metropolitan Statistical Area (MSA), Census Tract, Traffic Analysis District (TAD), and Traffic Analysis Zone (TAZ). Similar to the previous CTPPs, the 2006-2010 CTPP also consists of three parts. Part 1 is based on homes, such as number of resident workers. Part 2 is on workplaces, such as number of jobs. Part 3 provides detailed journey-to-work flow data, such as number of commuters between two census tracts by a specific mode

(e.g., drive alone, car pool, and others that include public transit, bicycle, walk and taxicab) and mean commuting time. No travel distances are reported.

As the core of Baton Rouge Metropolitan Area, East Baton Rouge Parish in Louisiana is selected as our study area. Other parishes in the metropolitan area are mostly rural. Parish is a county-equivalent unit in Louisiana. This research uses census tract as the zone unit for analysis since data at the census tract level is more reliable than at the TAZ level (http://data5.ctpp.transportation.org/ctpp/Browse/browsetables.aspx). There are 92 census tracts in the study area (excluding the airport tract with no records of any resident workers or jobs).

As mentioned previously, some of the records on journey-to-work time in the data set raise questions. For example, the reported mean travel time was 3.7 hours from tract 36.04 to tract 11.04 (with a corresponding estimated travel distance 5.4 miles), 2.8 hours from tract 22 to 26.01 (with an estimated travel distance 2.6 miles), and 2.9 hours from tract 1 to 16 (for an estimated travel distance of 4.5 miles), and so on.

In preparation of measuring network travel time and distance, we extracted the road network from the 2010 TIGER files from the U.S. Census Bureau (https://www.census.gov/geo/maps-data/data/tiger-line.html). Commuters were assumed to choose the shortest path by following the speed limits, and travel time and distance were estimated in ArcGIS. Research suggests that ArcGIS tend to underestimate travel time than what is actually experienced by travelers such as that retrieved from the Google Maps (Wang and Xu, 2011) and historical data (Jiang and Li, 2013). However, the two are highly correlated. For the purpose of comparison between actual and optimal commute lengths for analysis of excess commuting, it is adequate if both are based on estimated values as explained previously.

## Methodology

Excess commuting $T_w$ is the proportion of the average actual commute $T_a$ over the average required commute $T_r$, i.e.,

$$T_w = (T_a - T_r)/T_a \quad (1)$$

Figure 1 outlines the workflow of our analysis. We first replicate the existing approach of measuring excess commuting at the zonal (census tract) level, then use the Monte Carlo method to simulate individual locations of resident workers and employment as well as trips between them, and finally calibrate excess commuting at the individual level.

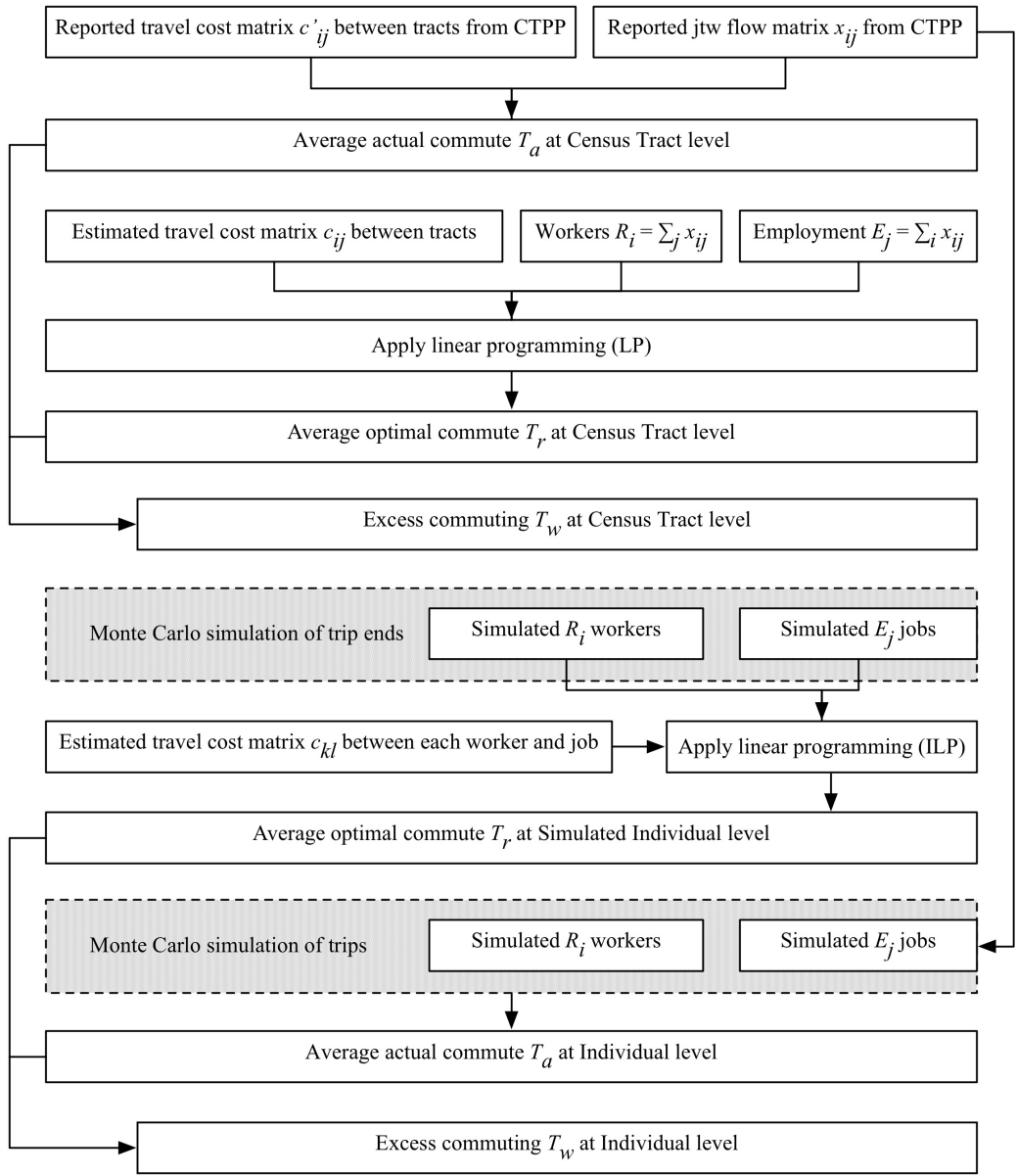

Figure 1. Workflow of measuring excess commuting at the zonal and simulated individual levels

Measuring Excess Commuting at the Census Tract Level

As stated previously, both travel time and distance are used to measure excess commuting. For demonstration of the methods, here only travel time is mentioned. We adopt the popular Linear Programming (LP) technique to derive optimal commute. Commonly referred to as the *transportation problem*, the problem is to solve the optimal journey-to-work flows between origin and destination zones in order to minimize the average travel time (Hitchcock, 1941; Horner and Murray, 2002; Taaffe et al., 1996), such as:

Minimize
$$T_r = \sum_i \sum_j (c_{ij} x_{ij}) / N \quad (2)$$

Subject to:
$$\sum_j x_{ij} = R_i \quad (3)$$
$$\sum_i x_{ij} = E_j \quad (4)$$
$$x_{ij} \geq 0 \quad (5)$$

where $c_{ij}$ is the commuting time from zone $i$ to zone $j$; $x_{ij}$ is the number of commuters living in zone $i$ and working in zone $j$; $R_i$ represents the number of commuters living in zone $i$; $E_j$ represents the number of commuters working in zone $j$; $N = \Sigma_i \Sigma_j x_{ij}$ is the total number of commuters.

Equation (2) defines the objective function of minimizing the average commuting time, which is subject to three constraints. Equation (3) ensures that all journey-to-work flows originating from one zone satisfy the total number of commuters living in that zone. Similarly, Equation (4) limits all journey-to-work flows ending in one zone to the total number of commuters working in that zone. Equation (5) restricts the journey-to-work flow matrix $x_{ij}$ to be non-negative values. The resulting $T_r$ returned by the objective function represents the average required commuting time, suggested by the spatial arrangement of homes and jobs. Comparing required commute $T_r$ to actual commute $T_a$, Equation (1) measures the excess commuting rate.

In implementation, intrazonal travel distance $c_{ii}$ is approximated as the radius of a circle with the same area size as a zone (Frost et al., 1998; Horner and Murray, 2002), and the corresponding intrazonal travel time is obtained by simply assuming a constant travel speed of 25 mph through the distance. The interzonal travel distance (time) $c_{ij}$ is calibrated between the centroids of census tracts by the shortest network path, and then modified by adding the intrazonal components (distance or time) at both the origin and destination zones.

Monte Carlo Simulations of Resident Workers and Jobs and Trips Between Them

To mitigate the effect of zonal scale, we begin with a Monte Carlo approach to simulate individual locations of resident workers and jobs. *Monte Carlo simulation* is a numerical analysis technique that uses repeated random sampling to obtain the distribution of an unknown probabilistic entity. It provides a powerful computational framework for spatial analysis, and has been used to disaggregate data in areal units to individual points. For example, Watanatada and Ben-Akiva (1979) used it to simulate representative individuals distributed in an urban area for travel demand analysis. Wegener (1985) designed a Monte Carlo based housing market model to analyze location decisions of industry and households, and corresponding migration and travel patterns. Luo et al. (2010) used it to randomly disaggregate cancer cases from the zip code level to census blocks in proportion to the age-race composition, and examined implications of spatial aggregation error in public health research. Gao et al. (2013) applied it to simulate resident and business distributions proportionally to mobile phone Erlang values and to predict traffic-flow in a road network.

Basically, a Monte Carlo method generates suitable random numbers of parameters or inputs to explore the behavior of a complex system or process. The random numbers generated follow a certain *probability distribution function* (PDF) that describes the occurrence probability of an event. Some common PDFs include: normal, uniform, and discrete distribution. This research uses it to simulate individual locations of resident workers and jobs that are proportional to their corresponding numbers within each census tract (i.e., following a discrete distribution). Specifically, the number of simulated trip origins is proportional to the number of resident workers in a tract, the number of trip destinations is proportional to the number of jobs there, and both are randomly distributed within the tract's boundary. In other words, denoting the total numbers of simulated and actual commuters by $n$ and $N$, respectively, and given the numbers of resident workers and employment in a tract $i$ from the CTPP as $R_i$ and $E_i$, the numbers of simulated individual workers and jobs in the tract $i$ are $(n/N)R_i$ and $(n/N)E_i$, respectively. The value of $n$ is selected by balancing accuracy and computational efficiency (i.e., larger numbers of simulated points and subsequently O-D trips improve accuracy but demand more computing power). For example, given the zonal level spatial patterns of resident workers and jobs in Figures 2(a) and 2(c), respectively, Figure 2(b) and 2(d) show corresponding simulated individual locations of resident workers and jobs.

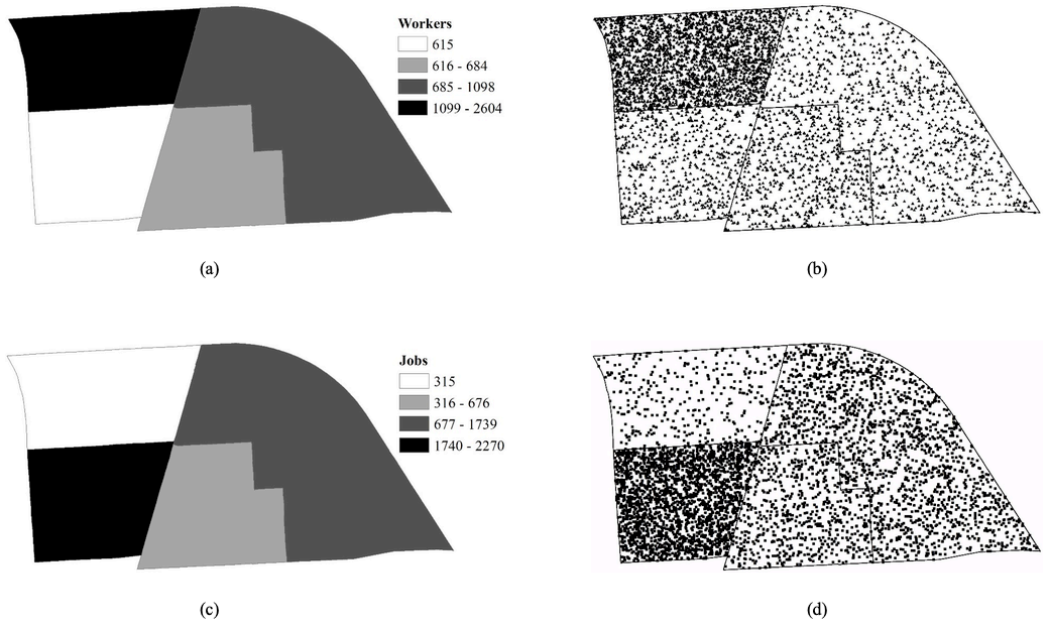

Figure 2. Spatial distribution of (a) resident workers in zones, (b) simulated resident workers, (c) jobs in zones, and (d) simulated jobs

The next step is to match a worker with a potential job to form the OD flow matrix. In preparing the OD travel time matrix for the optimization analysis, all possible trips from each of the resident worker locations to each of the job locations are calibrated.

For computing the average existing commute time, shown in Figure 1, we need to simulate the trips between individual locations of workers and jobs that are consistent with the actual zonal-level flows. Similarly, as the actual zonal-level flow from a residential worker tract $i$ to an employment tract $j$ (extracted from the CTPP) is $x_{ij}$, the simulated flow when aggregated at the zonal level should be $(n/N)x_{ij}$, i.e., proportional to the actual journey-to-work pattern. This is implemented by another Monte Carlo simulation process by utilizing the previously simulated points of resident workers (O) and jobs (D):

1) Randomly choosing an origin point $p(i)$ in a residential worker tract $i$,
2) Randomly choosing a destination point $q(j)$ in a job tract $j$,
3) Forming individual O-D trips between $p$ and $q$ and recording the trip length between them through the road network $c_{pq}$, and
4) Capping the number of trips between zones $i$ and $j$ at $(n/N)x_{ij}$.

See Hu and Wang (2015) for the program implementing the Monte Carlo Simulations.

Measuring Excess Commuting at the Simulated Individual Level

Similarly to the LP at the zonal level, the formulation of optimal commute at the individual level utilizes the Monte Carlo simulation of resident workers and jobs. Index the locations for simulated individual resident workers and jobs as $k$ and $l$, respectively. The total number of simulated workers is the same as that of simulated jobs, denoted by $n$. The optimization problem is

$$\text{Minimize} \quad \sum_{k=1}^{n}\sum_{l=1}^{n}(c_{kl}f_{kl})/n \quad (6)$$

Subject to:

$$\sum_{l=1}^{n} f_{kl} = 1 \quad (7)$$

$$\sum_{k=1}^{n} f_{kl} = 1 \quad (8)$$

$f_{kl} = 1$ when a trip from $k$ to $l$ is chosen, 0 otherwise (9)

where $c_{kl}$ is the estimated travel time from resident worker location $k$ to job location $l$, and $f_{kl}$ indicates whether a journey-to-work flow is chosen by the optimization algorithm ($= 1$ when chosen and 0 otherwise). The objective function (6) is to minimize the average commute time of $n$ simulated commuters. Equations (7) and (8) ensure that each worker can be assigned to one unique job and vice versa. Since the variable $f$ is a binary integer, it is an integer linear program (ILP) problem.

The result from the above ILP is the average minimal commuting, $T_r$. As explained previously, if the estimated travel time for a Monte Carlo simulated trip between two individual points $p$ and $q$ is $c_{pq}$, the average estimated commuting time $T_a$ in the simulated pattern is $T_a = \sum_{p,q} c_{pq}/n$. Since both the origins and destinations are individual points and do not involve any area configuration, both optimal commute and existing commute are estimated from the point-to-point OD trips. The approach is thus independent from the zonal effect or MAUP.

**Case Study in Baton Rouge**

Based on the 2006-2010 CTPP data in Baton Rouge, we first measure excess commuting at the census tract level based on White's (1988) LP approach, and then at the individual level based on our approach.

Excess Commuting at the Census Tract Level in Baton Rouge

Solving the zonal-level LP as defined in Equations (2)-(5), we obtain the optimal commuter flow matrix $x_{ij}$, and then the average minimum commute defined in Equation (2). The *average minimum commuting time* in the study area is 6.61 minutes, and the *average minimum commuting distance* is 3.44 miles. Note that the intrazonal travel times and distances are not zero here and vary with tract area sizes. As a result, the optimal commuting pattern has only 24.4% (i.e., 44,590/182,705) within-tract commute trips when travel time is used, and 19.5% (i.e., 35,597/182,705) within-tract trips when distance is used.

Figure 3 provides a visual comparison between actual and optimal commuter flows at the census tract level. Note that trips are substantially trimmed after optimization, and mostly are within tracts and between tracts in proximity.

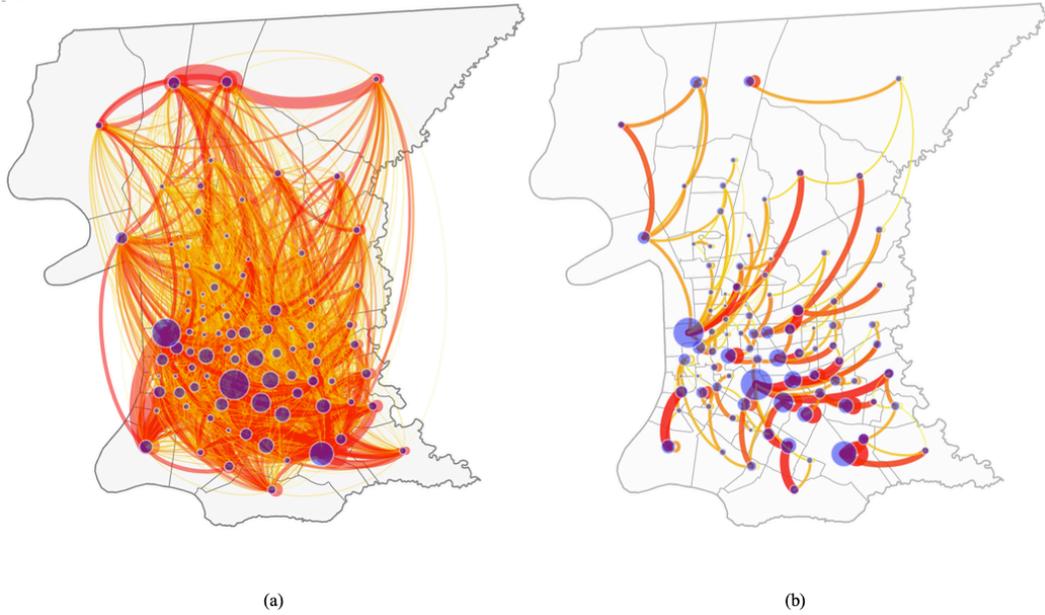

Figure 3. Tract-level commuting networks: (a) actual reported flow, and (b) optimal flow
[Line width is proportional to flow volume, and bubble size represents total throughput at a tract]

Based on reported journey-to-work flow volumes and corresponding mean travel time between tracts from the CTPP data, we obtained 19.26 minutes for the *average actual commuting time*. No mean travel distance is reported in the survey. As argued previously, reported travel time from a survey could include reporting errors and respondents might not be representative of the areas they reside. Furthermore, as the concept of excess commuting emphasizes the gap between actual and optimal commute lengths, it makes more sense to also estimate actual commute time (or distance) through network analysis since the optimal commute is an estimated measure. For comparison, we obtained the estimated travel time

(distance) for all commute flows, and then *average estimated commuting time* of 12.76 minutes and *average estimated commuting distance* of 7.42 miles.

With all the above results in place, excess commuting was calculated according to Equation (1) at the census tract level. Based on Table 1, excess commuting time was 65.68% by comparing actual reported time with optimal time, and dropped to 48.20% by comparing estimated time with optimal time. In terms of distance, excess commuting was 53.64%. The extent of excess commuting when using estimated time (48.20%) is more in line with that in distance (53.64%) as distance is also estimated from the road network.

Table 1. Summary of actual, optimal and excess commuting in Baton Rouge

|  |  |  | Average Commuting Time (min) | Average Commuting Distance (mile) |
|---|---|---|---|---|
| Census Tract Level | Actual | Reported | 19.26 | NA |
|  |  | Estimated | 12.76 | 7.42 |
|  | Optimal |  | 6.61 | 3.44 |
|  | Excess | Reported vs. Optimal | 65.68% | NA |
|  |  | Estimated vs. Optimal | 48.20% | 53.64% |
| Simulated Individual Level | Actual | Estimated | 12.65 | 7.63 |
|  | Optimal |  | 4.75 | 2.71 |
|  | Excess | Estimated vs. Optimal | 62.45% | 64.48% |

Note: NA means not available.

Excess Commuting at the Simulated Individual Level in Baton Rouge

To mitigate the scale effect in the zonal level analysis, we solved the integer linear programming (ILP) problem defined in Equations (6)-(9) for simulated individual commuters. As discussed previously, the total number of simulated commuters (or resident workers or jobs) $n$ by the Monte Carlo approach needs to be determined by balancing accuracy and computational efficiency. Based on a series of experiments with different sample sizes, we set the value of $n$ to be 3,565, which was limited by the computation of all possible OD cost matrix (i.e., 3,565*3,565 pairs) in ESRI ArcGIS 10.1 in a PC environment (i.e., Intel Core 2 Quad 2.4 GHz CPU with 4GB memory and 500GB hard disk). Possible strategies for remedy are discussed in next section. The ILP yielded *average minimum commuting time and distance* at the simulated individual level as 4.75 minutes and 2.71 miles, respectively. Both are significantly lower than the optimal commute time and distance obtained at the census tract level. This difference validates the impact of zonal effect on the measure of excess commuting as pointed out by Hamilton (1989).

Next we need to estimate travel time (distance) for the simulated trips that are consistent with the actual journey-to-work flow pattern. As discussed previously, the OD trips based on the Monte Carlo simulation are randomly paired individual locations of resident workers and jobs, but the total number of simulated commuters between two tracts $i$ and $j$ is capped proportionally to the actual journey-to-work volume such as $(n/N)x_{ij}$, where $n = 3,565$ and $N = 182,705$. By doing so, zonal-level trips are disaggregated into individual trips. Based on the simulation results, the *average estimated commuting time and distance* are 12.65 minutes and 7.63 miles, respectively. Both estimations are very close to those obtained at the census tract level. In other words, the disaggregation does not alter the estimated commuting amounts significantly, and the difference mainly lies in the minimum (required) commuting.

The visualization of networks at the simulated individual level would be too crowded to see any pattern. We aggregated both simulated flows and optimized flows into the census tract level, shown in Figure 4(a) and 4(b), respectively. Since the simulated flows at the individual level were intentionally designed to be proportional to the actual flows, the same pattern is observed between Figures 3(a) and 4(a), confirming that our simulation of trips worked well. In contrast to Figure 3(b) with a simpler pattern, Figure 4(b) shows the optimal commute flows at the census tract level that were aggregated back from the individual optimal pattern, which is far more complex. Individual workers are now free to swap houses for individual job locations, instead of being confined to a tract centroid for a group of workers (or a group of jobs). Therefore, the flexibility enables more choices in the optimized pairing between workers and jobs, and further brings down the total (average) minimum commuting. Figure 4b (aggregated from optimized individual trips) also shows far more interzonal trips than Figure 3b (directly optimized from zonal data) as individual workers are more likely to be paired with individual jobs in adjacent tracts instead of within the same tracts, and thus a more realistic optimization pattern.

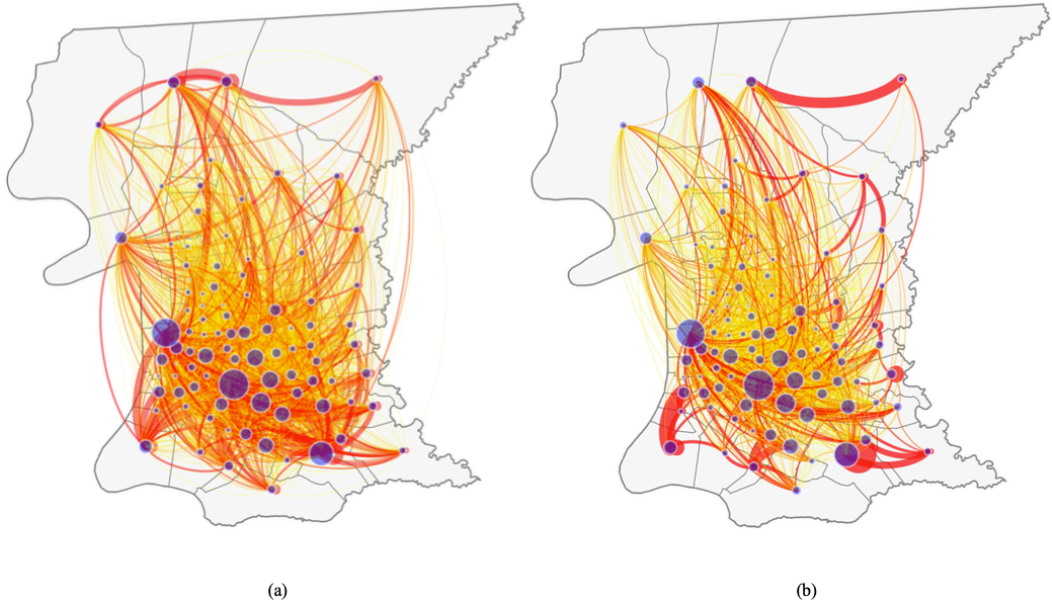

Figure 4. Aggregated individual-level commuting networks: (a) simulated flow and (b) optimal flow [Line width is proportional to flow volume, and bubble size represents total throughput at a tract]

Measured at the simulated individual level, we obtained 62.45% excess commuting time and 64.48% excess commuting distance. Both are higher than those estimated at the census tract level. The results are also reported in Table 1.

An issue regarding the *sensitivity of simulation sample size* merits some discussion here. The cap of $n$=3,565 simulated commuters was set due to our computation limitation in calibrating of the large $n \times n$ OD cost matrix for the ILP. Here we experimented with nine sample sizes from 820 to 3,565. Results for average optimal commute (time and distance) and corresponding excess commuting % are shown in Figure 5(a) and 5(b), respectively. As we increased the sample size, the average optimal commute, both in time and distance, tended to initially decrease and then stabilize after the sample size reached about 3,000. Even for smaller sample sizes (e.g., when $n$ increased from 820 to 1,282), the changes in both measures were minor (i.e., 0.13 minutes for optimal commuting time, and 0.1 miles for optimal commuting distance). For larger sample sizes (e.g., when $n$ increased from 3,256 to 3,565), the changes in the two measures were minimal (i.e., 0.004 minutes for time and 0.003 miles for distance). A similar trend can be said on the resulting percentages of excess commuting: the excess commuting time began to converge around 62% and the excess commuting distance tended to converge around 64% after the sample size reached 3,000. This confirms that the sample size of 3,565 commuters was a sound choice for the study area.

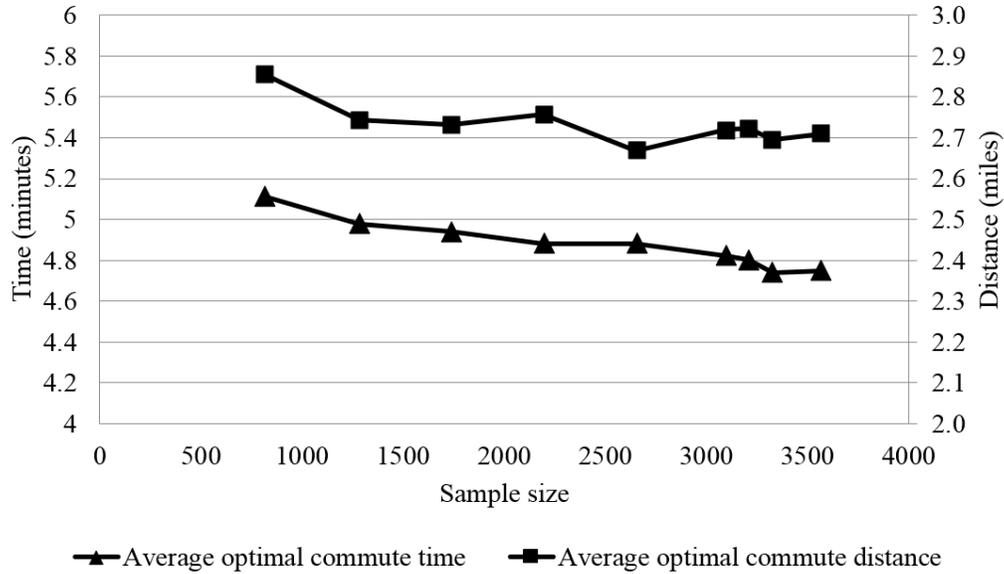

(a)

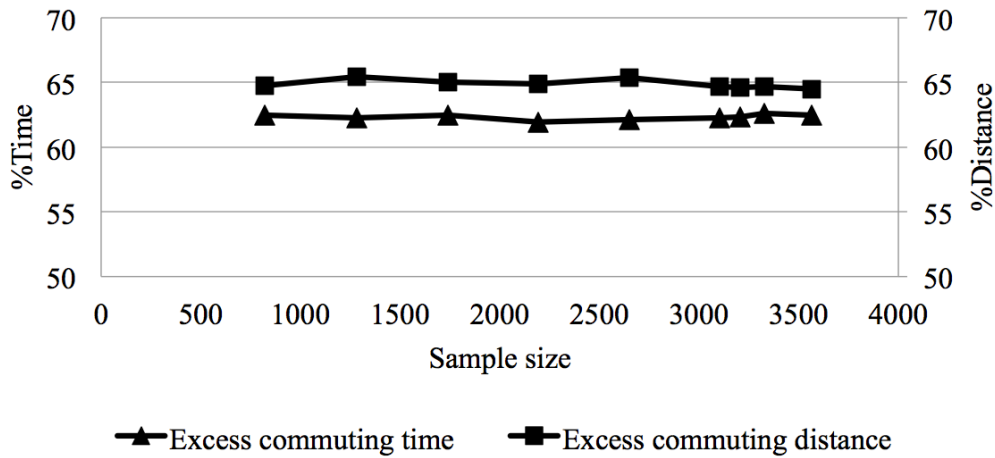

(b)

Figure 5. Sensitivity of simulation sample size: (a) optimal commute and (b) excess commuting %

**Discussion on Decomposing Excess Commuting**

Based on the results summarized in Table 1, Figure 6(a) illustrates various measures of average commuting time in the study area. The reported average time from survey stood at the highest value of 19.26 minutes. This may well be the amount of time experienced by commuters in general. However, the extent of excess commuting is derived when the actual commuting is compared to the minimal (optimal) commuting that is often based on estimation. The concept of excess commuting was initially proposed for the purpose of assessing the gap between what would be possible collectively for a city given its land use pattern and what individuals actually do. Therefore, it seems fairer to compare actual and optimal commute time by measuring both in estimated time from road network. Our average estimated commuting time (drove-alone) of 12.76 minutes was much shorter than the reported time of 19.26 minutes because the latter was affected by many factors such as commuters by slower modes, traffic congestion or even mental time that survey respondents might have included to account for parking and others. As the CTPP does not report commuting distance data, this issue is not relevant when distance is used to measure commuting length.

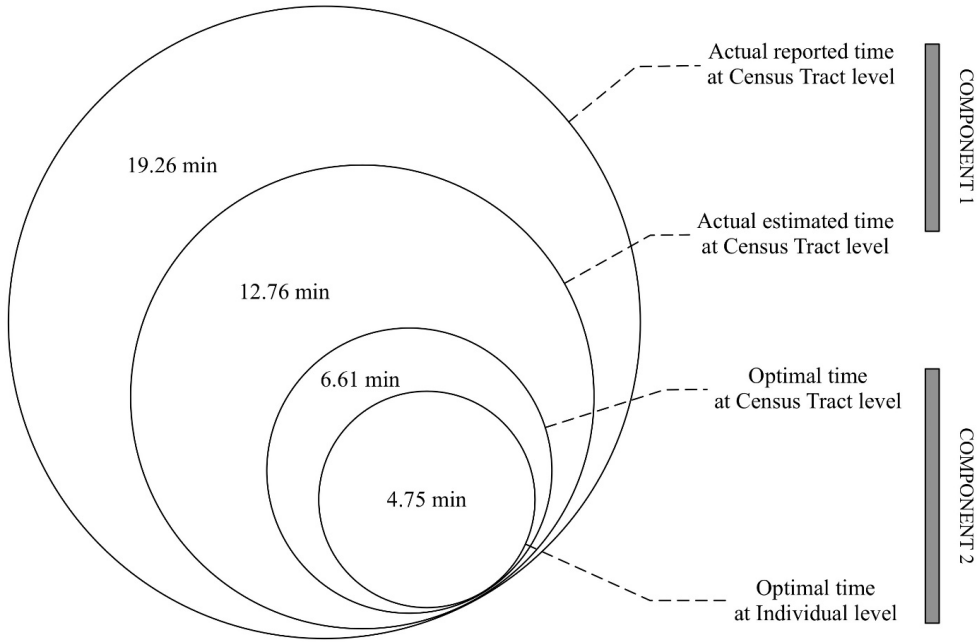

(a)

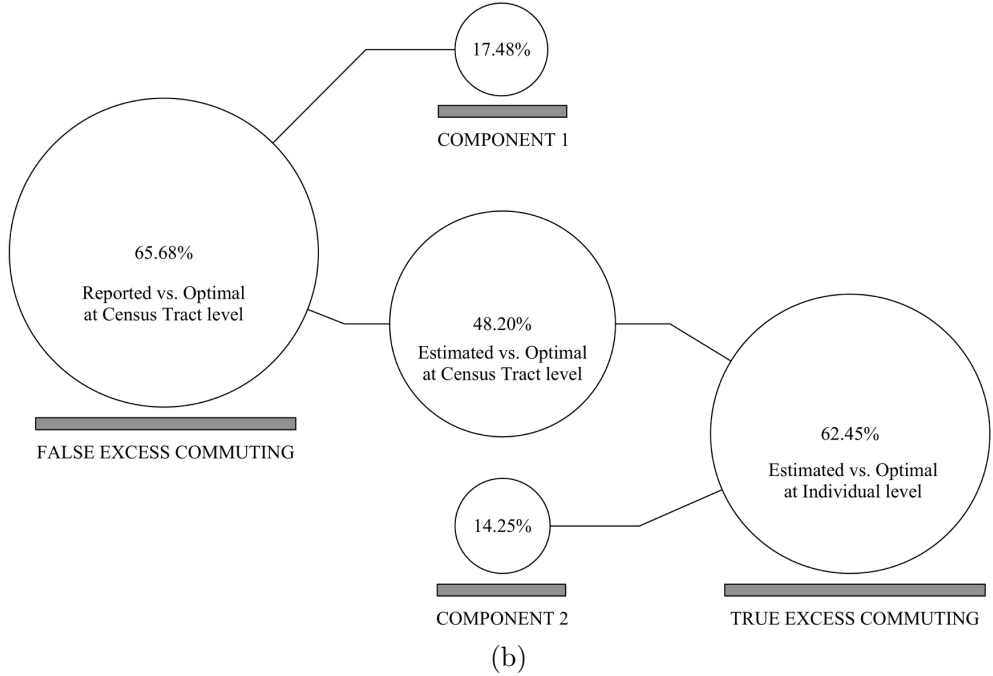

(b)

Figure 6. Decomposing excess commuting

The conventional LP approach at the zonal level yielded an average minimal (optimal) commuting time of 6.61 minutes. By simulating individual locations of resident workers and jobs and also the individual trips linking them, our research returned an average minimal commuting time of 4.75 minutes. When the zonal-level approach is used, resident workers and jobs are grouped together in zones (e.g., tract centroids), and the free swap of homes permitted by the optimization are between zones. In our approach at the simulated individual level, the optimization permits resident workers to freely swap homes with individual locations, leading to a large number of optimal commuting trips that are much shorter and interzonal. The individual level approach is not only more accurate for its sharper resolution in estimating trip lengths but also generates more realistic optimal commuting pattern.

Figure 6(b) further clarifies the two components of false estimation of excess commuting. When reported commuting time is used as a benchmark, the excess commuting stands as high as 65.68% at the zonal level. When estimated commuting time is used as a new benchmark, the excess commuting comes down to 48.20% at the zonal level. That is to say, using reported commuting time implies an *overestimation* of 17.48% excess commuting that is attributable to people using slower transport modes or getting caught in traffic, which a planning scenario of freely swapping house would not help. The 17.48% overestimation is termed "component 1" in miscalculation. By disaggregating zonal-level commuting patterns to simulated individuals, the average optimal commuting time is reduced and leads to a higher percentage of excess commuting (i.e., climbing back to 62.45% from 48.20%). That is to say, the zonal level approach causes an *underestimation* of excess commuting by 14.25%, termed "component 2" in miscalculation.

In summary, the reported commuting time inflates the excess commuting measure by 17.48%, and the zonal effect underestimates it by 14.25%. The two components bias the estimation in opposite directions and offset each other to a large extent. The traditional approach adopted in most literature yields 65.68% excess commuting, very close to 62.45% obtained in our approach. However, "two wrongs do not make a right." *The debate of excess commuting should not be about who gets the specific percentage close, rather about the scientific soundness in the means by which one reaches that percentage.*

When measured in distance, estimation error caused by component 1 is absent. The zonal level analysis generates 53.64% excess commuting. The simulated individual level analysis yields a lower average minimal distance, and thus increases the excess commuting to 64.48%, in line with the 62.45% excess commuting time.

This study also sheds lights on the choice of time or distance in measuring excess commuting. Some studies argue that travel time is a more appropriate estimate of travel cost and an important determinant factor of travel behavior (Buliung and Kanaroglou, 2002; Gordon et al., 1991; Wachs et al., 1993). In this study, difference in the results for time and distance is insignificant (when estimated time is used). The only complexity is whether to use reported commuting time as a benchmark for actual commute. If we define excessive time in using public transit or other slower modes and delayed time in congestion as "excess", reported time would be an adequate choice.

**Conclusions**

The concept of excess commuting is proposed to assess the gap between actual commute and minimum commute a city could achieve if residents were free to swap houses. In other words, it captures the potential for a city to reduce its overall commuting given its spatial arrangement of homes and jobs. Studies usually rely on survey data such as the CTPP to define actual commute time, and measure the optimal commute at an aggregate zonal level by linear programming (LP). This research argues that reported commute time by respondents tends to overestimate actual commute length as it includes travel time by slower transportation modes (e.g., public transits, bicycling, walking and others) and delayed time due to congestion. The zonal level analysis of optimal commute also suffers from the scale effect. Both contribute to bias in the measurement of excess commuting. This research uses a Monte Carlo approach to simulate trip ends, i.e., individual resident workers (O) and individual jobs (D) within census tracts, in order to mitigate the zonal effect. It also computes estimated commute distance and time for actual journey-to-work trips as a new benchmark for existing commuting. The resulting estimates of excess commuting are largely consistent between measures in time and distance. Based on the 2006-2010 Census for Transportation Planning Package (CTPP) data in Baton Rouge, the case study controls both sources of miscalculation for excess commuting and yields excess commuting of 62.45% in terms of time and 64.48% in terms of distance.

One limitation of our approach is the demand for computation power in Monte Carlo based simulation, which also limits the sample size of simulated commuters. Our simulated sample of 3,565 commuters works well in the study area and may be replicated successfully in cities of a similar size. However, challenges are conceivably bigger in larger metropolitan areas such as Los Angeles and Chicago. The sample size limitation in our experiment is resulted from neither the simulation of trip ends nor the simulation of trip distribution, but the extensive computation of network travel cost matrix between all possible trip ends in ArcGIS network analysis. One possible remedy is to compute the Euclidean distance matrix prior to the network-based OD cost matrix measure, and then apply an artificially large value for all long trips and only calibrate actual network distances for short trips. The reason is that the sole purpose of preparing the OD cost matrix is for the subsequent ILP optimization whose result usually does not include long trips. By doing so, the size of OD cost matrix is substantially reduced and frees us more computation power to increase the sample size of simulation in larger cities.

Openshaw, S., Taylor, P. J., 1979. A million or so correlation coefficients: three experiments on the modifiable areal unit problem. In: Wrigley, N., (Eds), *Statistical Applications in Spatial Sciences*. London: Pion, pp. 127-144.

Rodríguez, D. A., 2004. Spatial choices and excess commuting: a case study of bank tellers in Bogota, Colombia. *Journal of Transport Geography* 12(1), 49-61.

Scott, D. M., Kanaroglou, P. S., Anderson, W. P., 1997. Impacts of commuting efficiency on congestion and emissions: case of the Hamilton CMA, Canada. *Transportation Research Part D: Transport and Environment* 2(4), 245-257.

Small, K. A., Song, S., 1992. "Wasteful" commuting: a resolution. *Journal of Political Economy* 100(4), 888-898.

Suh, S. H., 1990. Wasteful commuting: an alternative approach. *Journal of Urban Economics* 28(3), 277-286.

Sultana, S., 2002. Job/housing imbalance and commuting time in the Atlanta metropolitan area: exploration of causes of longer commuting time. *Urban Geography* 23(8), 728-749.

Sultana, S., Weber, J., 2014. The nature of urban growth and the commuting transition: endless sprawl or a growth wave? *Urban Studies* 51(3), 544-576.

Taaffe, E. J., Gauthier, H. L., O'Kelly, M., 1996. Geography of Transportation, 2nd Edition. Upper Saddle River, NJ: Prentice Hall.

Thurston, L., Yezer, A. M., 1991. Testing the monocentric urban model: evidence based on wasteful commuting. *Real Estate Economics* 19(1), 41-51.

Wachs, M., Taylor, B. D., Levine, N., Ong, P., 1993. The changing commute: A case-study of the jobs-housing relationship over time. *Urban Studies* 30(10), 1711-1729.

Wang, F., 2001. Explaining intraurban variations of commuting by job proximity and workers' characteristics. *Environment and Planning B* 28(2), 169-182.

Wang, F., 2003. Job proximity and accessibility for workers of various wage groups. *Urban Geography* 24(3), 253-271.

Wang, F., Xu. Y., 2011. Estimating O-D matrix of travel time by Google Maps API: implementation, advantages and implications. *Annals of GIS* 17, 199-209.

Watanatada, T., Ben-Akiva, M., 1979. Forecasting urban travel demand for quick policy analysis with disaggregate choice models: A Monte Carlo simulation approach. *Transportation Research Part A* 13, 241-248.

Wegener, M., 1985. The Dortmund housing market model: a Monte Carlo simulation of a regional housing market. *Lecture Notes in Economics and Mathematical Systems* 239, 144-191.

White, M. J., 1988. Urban commuting journeys are not "wasteful". *Journal of Political Economy* 96(5), 1097-1110.

U.S. Census Bureau, 2012. U.S. Census Bureau, Statistical Abstract of the United States: 2012, Table 7,21